\documentclass[]{spie}  %>>> use for US letter paper
\usepackage[]{graphicx}

\title{Limiting noises in gravitational wave detectors: guidance from
their statistical properties.}

\author{Gabriela Gonz{\'a}lez
\skiplinehalf
Department of Physics and Astronomy \\
Louisiana State University \\
202 Nicholson Hall, Tower Drive \\
Baton Rouge, LA 70803
}

\begin{document} 
  \maketitle 

%%%%%%%%%%%%%%%%%%%%%%%%%%%%%%%%%%%%%%%%%%%%%%%%%%%%%%%%%%%%% 
\begin{abstract}
It is expected that interferometric gravitational wave detectors such
as LIGO \cite{Barish99} will be eventually limited by fundamental noise
sources like shot noise and Brownian motion, as well as by seismic
noise. In the commissioning process, other technical noise sources
(electronics noise, alignment fluctuations) limit the sensitivity and
are eliminated one by one. We propose here a way to correlate the
noise in the output of the gravitational wave detector with other
detector and environmental signals not through their linear transfer
functions (often unknown), but through their statistical
properties. This could prove useful for identifying the frequency
bands dominated by different noise sources in the final configuration,
and also to help the commissioning process.

\end{abstract}

%>>>> Include a list of keywords after the abstract 


%%%%%%%%%%%%%%%%%%%%%%%%%%%%%%%%%%%%%%%%%%%%%%%%%%%%%%%%%%%%%
\section{INTRODUCTION}
\label{sect:intro}  % \label{} allows reference to this section

It is expected that interferometric gravitational wave detectors such
as LIGO will be eventually limited by fundamental noise sources like
shot noise and Brownian motion, as well as by seismic noise. In the
commissioning process, other technical noise sources (electronics
noise, alignment fluctuations) limit the sensitivity and are
eliminated one by one. We propose here a way to correlate the noise in
the output of the gravitational wave detector with other detector and
environmental signals not through their linear transfer functions
(often unknown), but through their statistical properties. This method
was very useful in distinguishing the frequency band where Brownian
motion was dominant, from other bands where seismic noise was
dominant, in Ref.\citenum{Gonzalez95}. This method could then prove
useful for identifying the frequency bands dominated by different
noise sources in the final configuration of gravitational wave
detectors, and also to help the commissioning process. we present in
this article some preliminary results obtained with data from the LIGO
first Science Run (August 23-September 9, 2002)\cite{S1}. 

\section{NOISE STATISTICS IN FREQUENCY DOMAIN}

For some special classes of noise, statistical properties in the
frequency domain are well defined in standard textbooks
\cite{BendatPiersol86}. Assume we are given a time series $x(t)$
digitized with a sampling time $\Delta t=1/f_s$, and measured over a
time interval $T$: $x_n=x(t_n)$, $t_n={0, \Delta t, ...N\Delta
t=T}$. We can then calculate an estimate of its power spectral density
by averaging over K=N/M power spectral densities ${S_x}_i(f)$,
i=1...K, calculated for shorter time intervals of duration $T/K$. If
the process is random and stationary, each time series $x_i(t_n)$ with
M points $n=(i-1)N/K+1 ... iN/K$ is a representative of an ensemble we
can use to calculate means and variances at each frequency
$f_k=2k/f_s$.

$${S_x}_i(f_k)=\frac{2\Delta t}{M}
\left(\sum_{n,m=(i-1)M+1}^{i M} x_n x_m e^{-2i\pi(k-1)(n-m)/M}\right)$$

$$S_x(f_k)=(1/K)\sum_{i=1}^{K}{{S_x}_i(f_k)}.$$ 

We can also also calculate the standard deviation of each frequency
bin:

$$E_x(f_k)=(1/N)\sum \left({S_x}_i(f_k)-S_x(f_k)\right)^2$$

If the signal $x(t)$ is a stationary, Gaussian process, then the ratio
$R(f)=E(f)/S(f)$ has a mean of one with a variance equal to
$1/\sqrt{N}$. If the signal is a sine wave, then $E(f)$ will tend to
zero. If the signal is a random, stationary background with impulsive
transients (``glitches'') added, $E(f)$ will in general will be larger
than one. 

The output signal in gravitational wave detectors, as in many other
complex experiments, is a sum of many different noise components. The
statistical properties of the time series $x(t)$ , will strongly
depend on the spectral shape of the signal. Detection of signals in
the noise are most often done in the frequency domain for this reason;
if they are done in the time domain, it is usually after applying
filters to select signals in a frequency band of interest, and also
whitening the data.

A very important problem, different than signal detection, is
identifying the contributions of different noise sources to the final
output signal. This is done while ``commissioning'' instruments such
as gravitational wave detectors, and is the first step to help
eliminate unwanted or unexpected sources of noise, dominant over the
expected ultimate limiting sources of noise. In general, the
diagnostics is done by assuming known linear processes from sources
whose spectral density can be measured independently (electronics
noise and seismic noise, for example). Then, a comparison is made in
the frequency domain, of the expected noise (source noise through a
linear transformation), with the measured output spectral density. In
general, different noise sources will be dominant in different
frequency bands. This method is efficient only when the transfer
functions are well known. However, many times the transfer function is
unknown, and difficult or impossible to measure.

The statistical properties of the signals of the dominant sources of
noise should be preserved by linear transformations, and thus the
features of the function $R(f)=E(f)/S(f)$ (i.e., whether it is larger
than one, close to one, or smaller than one) provide an opportunity to
recognize the sources of noise even without the knowledge of the
transfer function.  We provide some examples of possible uses for the
noise diagnostics in signals from the LIGO gravitational wave
detectors. 

\section{Noise in Gravitational Wave Detectors}

In LIGO gravitational wave detectors, the signal which is later
calibrated and interpreted as the gravitational wave signal is the
error signal in a feedback loop, controlling a demodulated signal of a
photocurrent detected at the antisymmetric port of the interferometer. 

\begin{figure}
   \begin{center}
   \begin{tabular}{c}
   \includegraphics[width=6in]{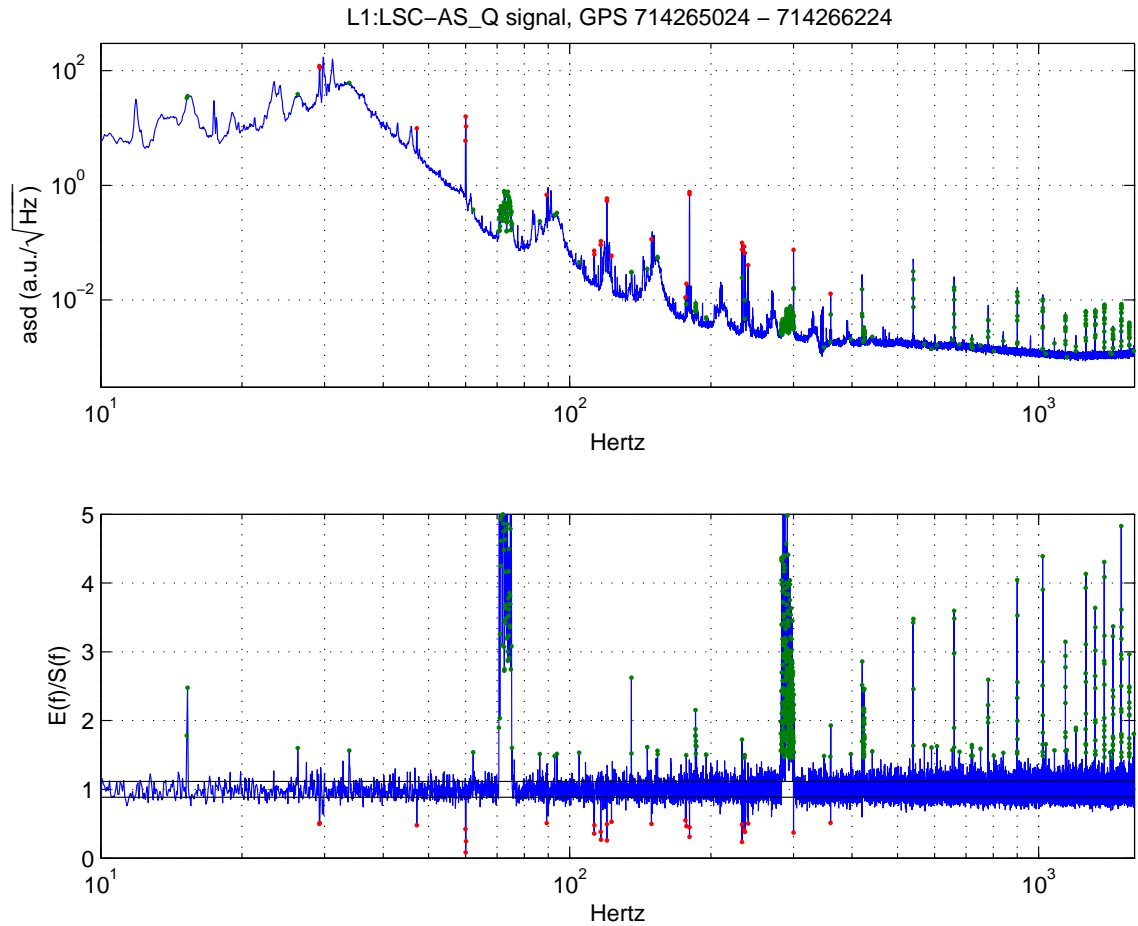}
   \end{tabular}
   \end{center}
   \caption[example] 
%>>>> use \label inside caption to get Fig. number with \ref{}
   { \label{f:ASQpsdstats} Top: Amplitude Spectral Density of the
output of the LIGO Livingston Gravitational Wave Signal (the units of
the raw signal are arbitrary). Bottom: The ratio of variance to mean
in each frequency bin. The points indicate the bins outside a
2-$\sigma$ interval of a random, stationary noise source. The solid
lines indicate 1-$\sigma$ deviation from unity.}
   \end{figure}

We present in Fig.~\ref{f:ASQpsdstats} the power spectral density of
this signal, taken during 1216 seconds of the continuous operation in
the first Science Run of the LIGO Science Collaboration. We have used
a sampling frequency of $f_s=$ 4096 Hz, took K=76 measurements of
power spectra, each for a segment 16 seconds long. The frequency bins
are 62.5 mHz wide.

We show in Fig.~\ref{f:ASQpsdstats} the calculation of the function
$R(f)=E(f)/S(f)$.  For random, stationary noise we expect this
function to be close to one, within $1/\sqrt{N}$, where $N$ is the
number of averages. We show these error bars for this particular
measurement, taken with N=16 averages of $E(f)$ and $S(f)$ calculated
for 16 seconds intervals (for a total time of 1200 seconds). A large
fraction of the points have the statistical character of random,
stationary noise: 94\% of the frequency bins have $R(f)$ within 23\%
(or $2/\sqrt{K}$ of the expected value of one. However, many other
frequency bins have a value of $R(f)$ significantly larger or smaller
than one. Indicated in Fig\ref{f:ASQpsdstats} are the points with
$R(f)$ outside a 3-$\sigma$ interval from unity. The largest values of
$R$ are mostly concentrated in two frequency bands: 70-76 Hz and
280-300 Hz, corresponding to two small broad peaks in the amplitude
spectral density. The frequencies with low values of $R$ (indicating
possible periodic signals) are all narrow lines in the spectrum,
mostly low harmonics of the 60 Hz AC line. Interestingly, high
harmonics of the power line frequency ($\geq 420$ Hz) have {\em high}
values of $R(f)$, opposed to the low harmonics ( $\leq 360$ Hz).

This simple calculation on a single signal shows already qualitative
differences in the statistics of different frequency bands, showing
that this strategy may prove very useful. Notice that the features of
the function $R(f)$ are very different than the features in the
amplitude spectral density. We now discuss in detail some further
analysis of different frequency bands.

%-------------
%-------------

\subsection{Stationary frequency bands}

\begin{figure}
   \begin{center}
   \begin{tabular}{c}
   \includegraphics[width=6in]{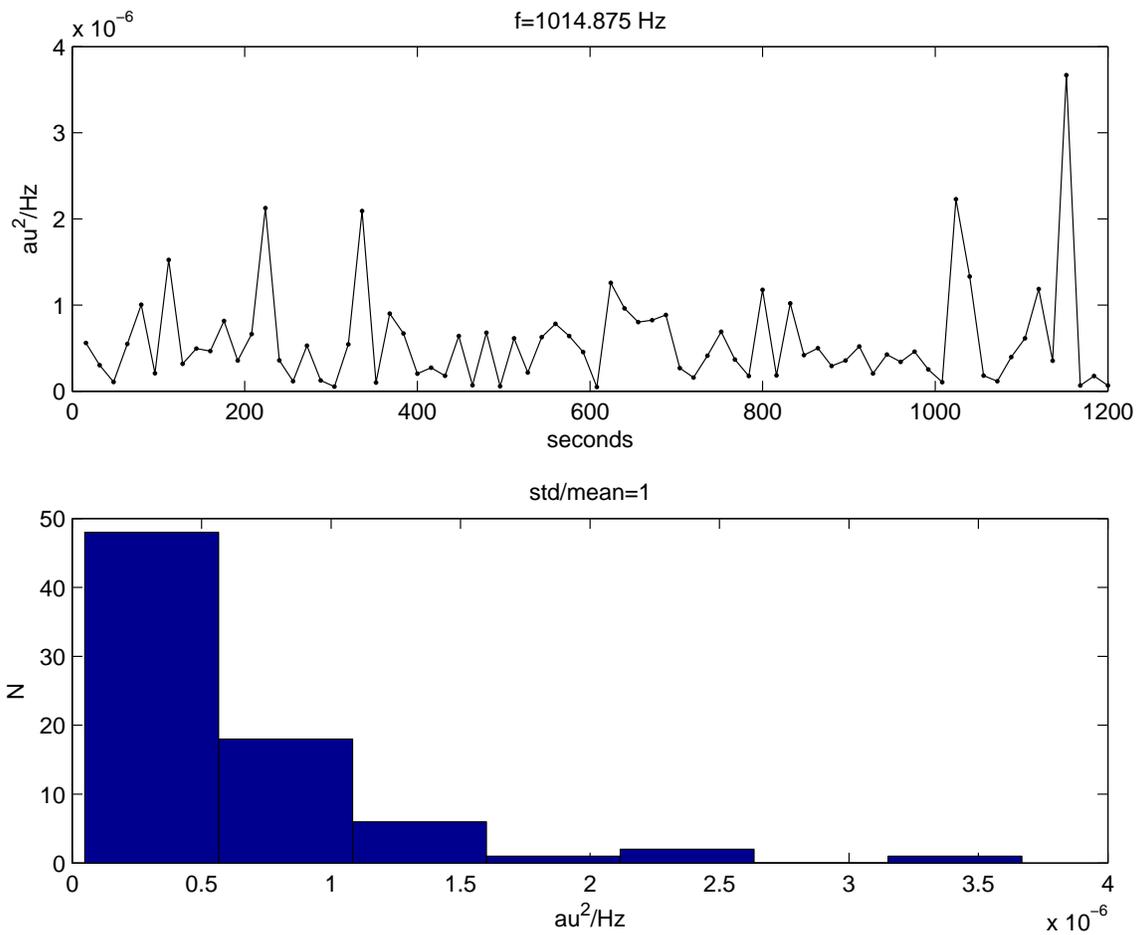}
   \end{tabular}
   \end{center}
   \caption[example] 
%>>>> use \label inside caption to get Fig. number with \ref{}
   { \label{f:goodstatex} Top: time history of the power in the
frequency bin indicated. Bottom: histogram of the values shown in
top. The numerical ratio of variance to mean is 1+$7\times10^{-8}$
(but of course the statistical error in the estimates of the mean and
variance are larger, given by $1/\sqrt{K}=0.1$).}
   \end{figure}

Most of the frequency bins shown in Fig.~\ref{f:ASQpsdstats} have
statistics consistent with random, stationary noise. An example of the
time history and histogram of the power in a single frequency bin is
shown in Fig.\ref{f:goodstatex}, for a frequency near 1kHz. We expect
most ultimately limiting noise sources (Brownian motion,shot noise)
and some of technical noise sources (electronic noise) to be random
and stationary. This test, then, does not provide a particular mean to
distinguish different noise sources {\em if} they are all random and
stationary. As we see in the next examples, this is very often not the
case.

%-------------
%-------------

\subsection{Two noisy frequency bands}

As we noticed in Fig.~\ref{f:ASQpsdstats}, there are two broad
frequency bands which are strongly non-stationary, [70-76] Hz and
[280-300] Hz, corresponding to two small broad peaks in the amplitude
spectral density. Notice that the frequency resolution in the analysis
(63 mHz) allows us to distinguish the properties of the 300 Hz
frequency bin, with R=0.37 (indicating periodicity), from its two
neighboring bins, with R=0.9 (indicating random stationary noise),
and from lower frequency bins: the bin at 299.9375 Hz has R=2.1,
indicating strong non-stationary behavior.

\begin{figure}
   \begin{center}
   \begin{tabular}{c}
   \includegraphics[width=6in]{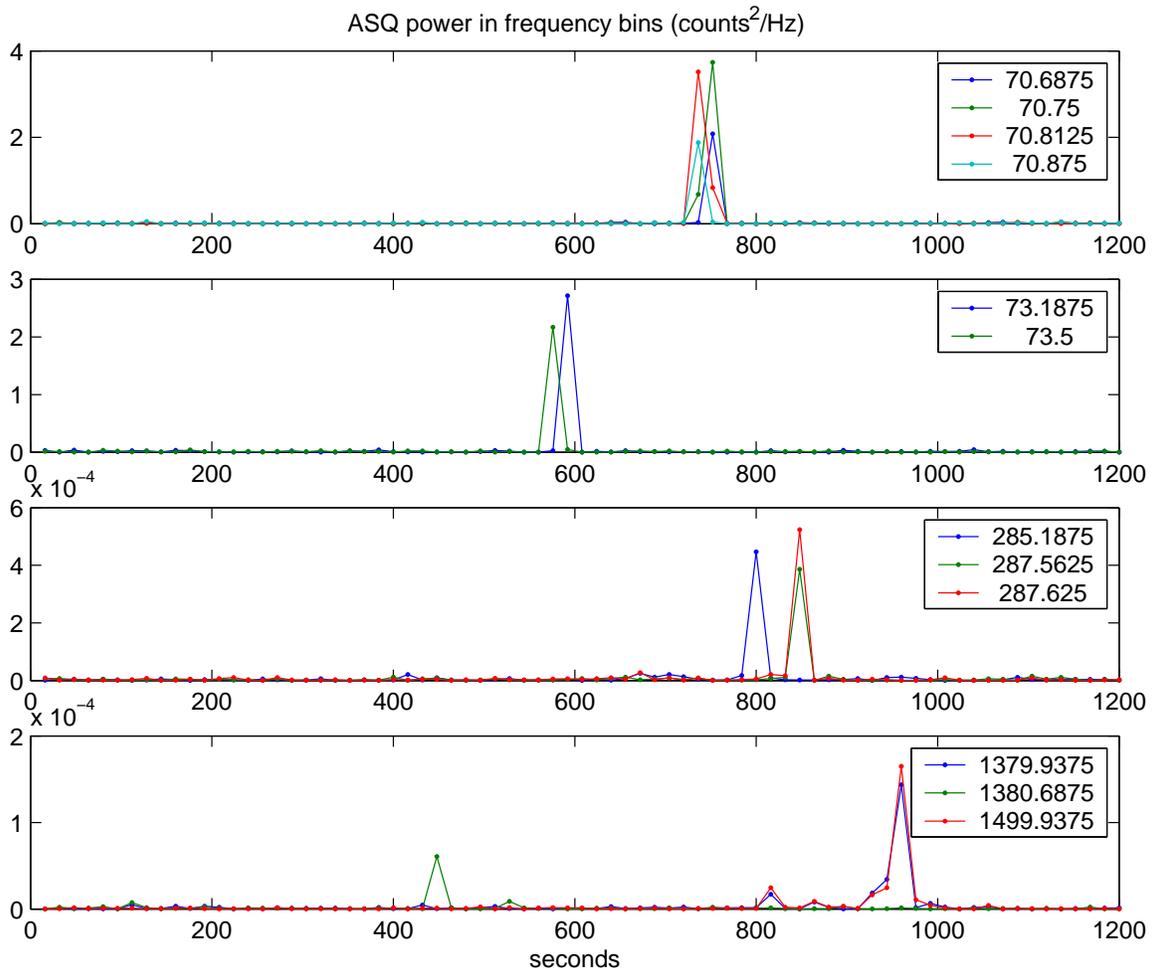}
   \end{tabular}
   \end{center}
   \caption[example] 
%>>>> use \label inside caption to get Fig. number with \ref{}
   { \label{f:psdvstime} Time histories of power in different
frequency bins that show a very non-stationary behavior, indicated by
a ratio of variance to mean larger than 2.}
   \end{figure}

If we plot the power in each frequency bin as a function of time, for
each 16 seconds time segment used in the ensemble, we notice that
usually the culprit for the non-stationary behavior is one or two
points with a large excess power, or a ``glitch''. These glitches
happen, however, at different times in the different frequency bins,
as shown in Fig.~\ref{f:psdvstime}. Since the glitches are
concentrated in definite frequency bands, it is likely that the source
of non-stationarity in each band is common; however it it clear that
the source produces recurrent glitches in slightly different frequency
bins. This particular feature could be fruitfully used to recognize
the characteristics in another signal, to help the identification of
the noise source.

%-------------
%-------------

\subsection{Power lines: two different sources?}

\begin{figure}
   \begin{center}
   \begin{tabular}{c}
   \includegraphics[width=6in]{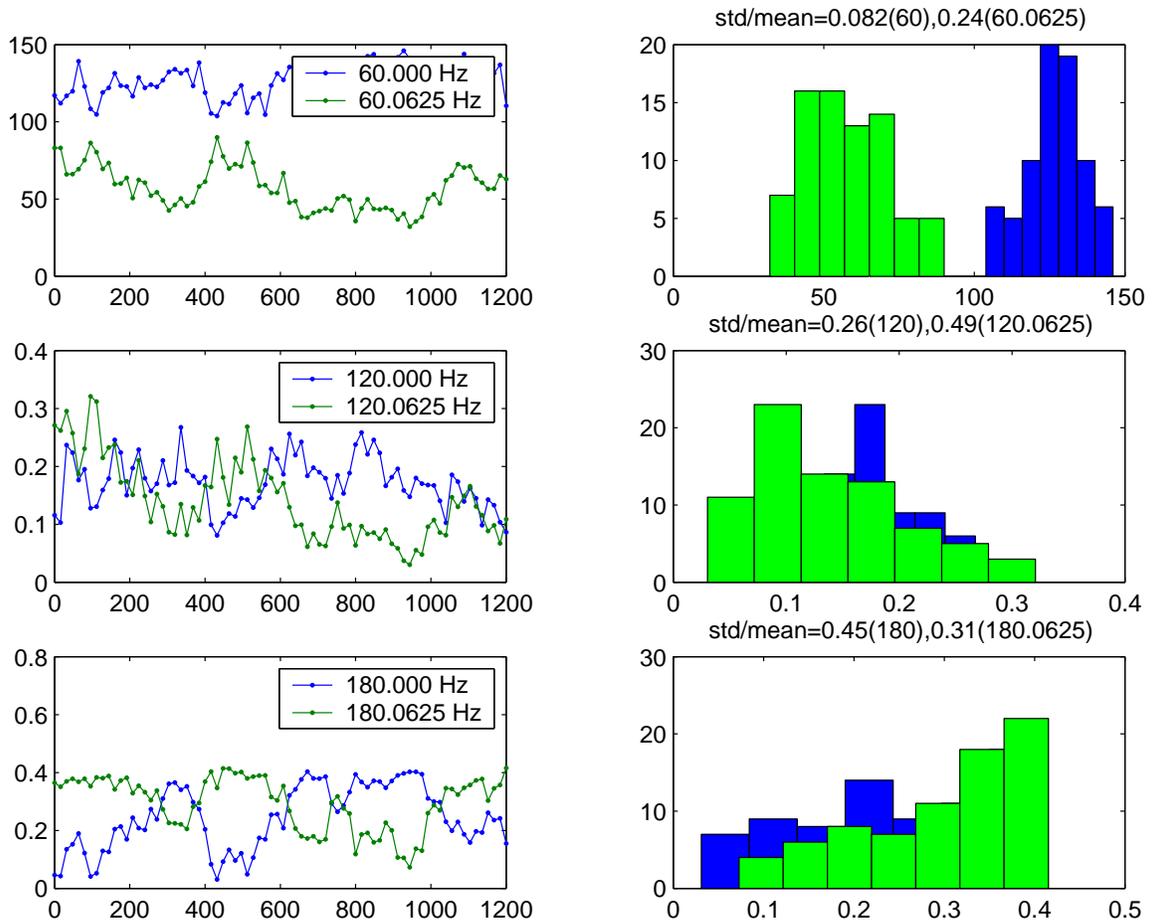}
   \end{tabular}
   \end{center}
   \caption[example] 
%>>>> use \label inside caption to get Fig. number with \ref{}
   { \label{f:hist60Hz} Time histories of power in frequency bins near
harmonics of 60 Hz, and corresponding histograms.}
   \end{figure}

The harmonic frequencies of the AC power line, 60 Hz, are a special
set. The low number harmonics have stationary, periodic structure
rather than random, with a very low R, as shown in
Fin.\ref{f:hist60Hz}: the power in the frequency bin has a mean
significantly different from zero, and the histogram approximates a
Gaussian away from zero rather than a Rayleigh distribution. This
features disappears gradually for higher harmonics, where a
non-stationary behavior is present, with the maximum power oscillating
between neighboring frequency bins: this is typical of beating of
lines, where some lines are from th AC power and some from driven
circuits or machinery which may lag the driving frequency. 

%-------------
%-------------

%\subsection{Violin modes}

\section{Conclusions}

We have shown that the study of statistical properties of signals in
interferometric wave detectors is very useful to distinguish dominant
sources of noise in different frequency bins. This may prove a
powerful technique when analyzing progress and differences in the
commissioning of interferometric gravitational wave detectors. It is
already being used as a graphical monitor in the LIGO Science runs
\cite{RayleighMon}, as a qualitative measure of the data quality. We
hope to use this method for more quantitative measures of the data
quality, but more importantly, to find the sources of the
non-stationary noise in the gravitational wave output.

%%%%%%%%%%%%%%%%%%%%%%%%%%%%%%%%%%%%%%%%%%%%%%%%%%%%%%%%%%%%%
\acknowledgments     %>>>> equivalent to \section*{ACKNOWLEDGMENTS}       
 
The author acknowledges many useful discussions about the use of
statistical methods with P. R. Saulson, L. S. Finn and P. Sutton. The
data used was taken during LIGO's first Science Run, which was run by
the LIGO Science Collaboration and the LIGO Laboratory. This work was
supported by Louisiana State University, and by the National Science
Foundation grant PHY-9870032.

%%%%%%%%%%%%%%%%%%%%%%%%%%%%%%%%%%%%%%%%%%%%%%%%%%%%%%%%%%%%%
%%%%% References %%%%%

\bibliography{GWNoiseStats}   %>>>> bibliography data in report.bib
\bibliographystyle{spiebib}   %>>>> makes bibtex use spiebib.bst

\end{document}